\documentclass[11pt]{article}

\usepackage{natbib}

\usepackage{setspace}

\usepackage[top=1in, bottom=1in, left=1in, right=1in]{geometry}
\usepackage{rotating}
\usepackage[nolists]{endfloat}
\usepackage{amsmath}
\usepackage{amssymb}
\usepackage{graphicx}

\title{Estimation for general birth-death processes}

\author{Forrest W. Crawford$^1$, Vladimir N. Minin$^2$, and Marc A. Suchard$^{3,4,5}$ }

\date{Typeset \today}

\newcommand{\od}[2]{\frac{\text{d} #1}{\text{d} #2}}

\newcommand{\dx}[1]{\ \text{d} #1}
\newcommand{\E}{\mathbb{E}}

\newcommand{\btheta}{\boldsymbol{\theta}}
\newcommand{\Y}{\mathbf{Y}}
\newcommand{\z}{\mathbf{z}}
\newcommand{\xmin}{x_\text{min}}
\newcommand{\argmax}[1]{\underset{#1}{\operatorname{argmax}}\ }

\begin{document}

%\twocolumn[
  %\begin{@twocolumnfalse}

   \maketitle

\begin{footnotesize}
	%\hspace{-1.8cm}
\begin{minipage}[t]{2.1in}
  \noindent 1. Department of Biomathematics\\
  University of California Los Angeles\\
  CHS AV-611\\
  Los Angeles, CA 90095-1766 USA \\
  fcrawford@ucla.edu 
\end{minipage}
%\hspace{0.02in}
\begin{minipage}[t]{2.1in}
2. Department of Statistics \\
University of Washington \\
Padelford Hall C-315, Box 354322 \\
Seattle, WA 98195-4322 USA \\
vminin@u.washington.edu
\end{minipage}
%\hspace{0.02in}
\begin{minipage}[t]{2.1in}
  3. Departments of Biomathematics, \\
  4. Human Genetics, and \\
  5. Biostatistics \\
  University of California Los Angeles\\
  6558 Gonda Building, \\
  Los Angeles, CA 90095-1766 USA\\
  msuchard@ucla.edu
\end{minipage}
\end{footnotesize}
		
\doublespacing

\begin{abstract}
\noindent Birth-death processes (BDPs) are continuous-time Markov chains that track the number of ``particles'' in a system over time.  While widely used in population biology, genetics and ecology, statistical inference of the instantaneous particle birth and death rates remains largely limited to restrictive linear BDPs in which per-particle birth and death rates are constant.  Researchers often observe the number of particles at discrete times, necessitating data augmentation procedures such as expectation-maximization (EM) to find maximum likelihood estimates.  The E-step in the EM algorithm is available in closed-form for some linear BDPs, but otherwise previous work has resorted to approximation or simulation.  Remarkably, the E-step conditional expectations can also be expressed as convolutions of computable transition probabilities for any general BDP with arbitrary rates.  This important observation, along with a convenient continued fraction representation of the Laplace transforms of the transition probabilities, allows novel and efficient computation of the conditional expectations for all BDPs, eliminating the need for approximation or costly simulation.  We use this insight to derive EM algorithms that yield maximum likelihood estimation for general BDPs characterized by various rate models, including generalized linear models.  We show that our Laplace convolution technique outperforms competing methods when available and demonstrate a technique to accelerate EM algorithm convergence.  Finally, we validate our approach using synthetic data and then apply our methods to estimation of mutation parameters in microsatellite evolution.
\end{abstract}

\vspace{0.3cm}


\vspace{0.3cm}

  %\end{@twocolumnfalse}
  %]

%%%%%%%%%%%%%%%%%%%%%%%%%%%%%%%%%%%%%%%%%%%%%%%%%%%
%\begin{large}

\section{Introduction}

A birth-death process (BDP) is a continuous-time Markov chain that models a non-negative integer number of particles in a system \citep{Feller1971Introduction}.  The state of the system at a given time is the number of particles in existence.  At any moment in time, one of the particles may ``give birth'' to a new particle, increasing the count by one, or one particle may ``die'', decreasing the count by one.  BDPs are popular modeling tools in a wide variety of quantitative disciplines, such as population biology, genetics, and ecology \citep{Thorne1991Evolutionary,Krone1997Ancestral,Novozhilov2006Biological}.  For example, BDPs can characterize epidemic dynamics, \citep{Bailey1964Elements,Andersson2000Stochastic}, speciation and extinction \citep{Nee1994Reconstructed,Nee2006Birth}, evolution of gene families \citep{Cotton2005Rates,Demuth2006Evolution}, and the insertion and deletion events for probabilistic alignment of DNA sequences \citep{Thorne1991Evolutionary,Holmes2001Evolutionary}.

Traditionally, most modeling applications have used the ``simple linear'' BDP with constant per-particle birth and death rates, which arises from an assumption of independence among particles and no background birth and death rates.  When individual birth and death rates instead depend on the size of the population as a whole, the model is called a ``general'' BDP. 
Previous statistical estimation in BDPs has focused mainly on estimating the constant per-particle birth and death rates of the simple linear BDP based on observations of the number of particles over time.
However, the simple linear BDP is often unrealistic, and nonlinear dependence of the birth and death rates on the current number of particles provides the means to model more sophisticated and realistic patterns of stochastic population dynamics in a wide variety of biological disciplines. %\citep{Novozhilov2006Biological}.  
For example, populations sometimes exhibit logistic-like growth as their number approaches the carrying capacity of their environment \citep{Tan1991Stochastic}.  In genetic models, the rate of new offspring carrying an allele often depends on the proportions of both individuals already carrying the allele and those who do not \citep{Moran1958Random}.  In coalescent theory, the rate of coalescence changes with the square of the number of lineages \citep{Kingman1982Genealogy}.  In addition, researchers may wish to assess the influence of covariates on birth and death rates by fitting a regression model \citep{Kalbfleisch1985Analysis,Liu2007Analysis}.

Progress in estimating birth and death rates in BDPs has also typically been limited to continuous observation of the process \citep{Moran1951Estimation,Moran1953Estimation,Anscombe1953Sequential,Darwin1956Behaviour,Wolff1965Problems,Reynolds1973Estimating,Keiding1975Maximum}.  However, in practice researchers may observe data from BDPs only at discrete times through longitudinal observations.
Estimating transition rates in continuous-time Markov processes using discrete observations is difficult since the state path between observations is not observed.  Furthermore, direct analytic maximization of the likelihood  for general BDPs remains infeasible for partially observed samples since the likelihood usually cannot be written in closed-form.  Despite these challenges, several researchers have made progress in estimating parameters of the simple linear BDP under discrete observation \citep{Keiding1974Estimation,Thorne1991Evolutionary,Holmes2001Evolutionary,Rosenberg2003Estimating,Dauxois2004Bayesian}. However, none of these developments provides a robust method to find exact maximum likelihood estimates (MLEs) of parameters in discretely observed general BDPs with arbitrary birth and death rates.

A major insight comes from the fact that the likelihood of the continuously observed process has a simple form which easily yields expressions for estimation of rate parameters.  This fact is the basis for expectation-maximization (EM) algorithms for maximum likelihood estimation in missing data problems \citep{Dempster1977Maximum}.  In finite state-space Markov chains, the relevant conditional expectations (the E-step of the EM algorithm) can often be computed efficiently, and several researchers have derived EM algorithms for estimating transition rates in this context \citep{Lange1995Gradient,Holmes2002Expectation,Hobolth2005Statistical,Bladt2005Statistical,Metzner2007Generator}.  Unfortunately, finding these conditional expectations for general BDPs poses challenges since the joint distribution of the states and waiting times (or its generating function) is usually not available in closed-form.  Notably, \citet{Holmes2001Evolutionary,Holmes2002Expectation} and \citet{Doss2010Great} are able to find analytic expressions or numerical approximations for these expectations in EM algorithms for certain BDPs whose rates depend linearly on the current number of particles.
While these developments are promising, there remains a great need for estimation techniques that can be applied to more sophisticated BDPs under a variety of sampling scenarios.  Indeed, more complex and realistic models like those reviewed by \citet{Novozhilov2006Biological} may be of little use to applied researchers if no practical method exists to estimate their parameters.

Here we seek to fill this apparent void by providing a framework for deriving EM algorithms for estimating rate parameters of a general BDP.  We first formally define the general BDP and give an exact expression for the Laplace transform of the transition probabilities in the form of a continued fraction.  We then give the likelihood for continuously-observed BDPs and outline the EM algorithm.  Next, we describe a novel method to efficiently compute the expectations of the E-step for BDPs with arbitrary rates.  Since these expectations are convolutions of transition probabilities, we perform the convolution in the Laplace domain, and then invert the Laplace transformed expressions to obtain the desired conditional expectation. This technique obviates the costly numerical integration or repeated simulation that has plagued previous approaches.  We provide examples of the maximization step for several different classes of BDPs and demonstrate a technique for accelerating convergence of the EM algorithm.  We show that our method is faster than competing techniques and validate it using simulated data.  Finally, we conclude with an application that analyzes microsatellite evolution and answers an open question in evolutionary genomics.

%%%%%%%%%%%%%%%%%%%%%%%%%%%%%%%%%%%%%%%%%%%%%%%%%%
\section{General BDPs and their EM algorithms}

\subsection{Formal description and transition probabilities}

\label{sec:transprobs}

Consider a general BDP $X(\tau)$ counting the number of particles $k$ in existence at times $\tau\geq 0$.  From state $X(\tau)=k$, transitions to state $k+1$ happen with instantaneous rate $\lambda_k$, and transitions to state $k-1$ happen with instantaneous rate $\mu_k$.  The transition rates $\lambda_k$ and $\mu_k$ may depend on $k$ but are time-homogeneous.  As we show below, it is often necessary to evaluate finite-time transition probabilities to derive efficient EM algorithms for estimation of arbitrary birth and death rates in general BDPs.  This proves useful both in completing the E-step of the EM algorithm and in computing incomplete data likelihoods for validation of our EM estimates.  For a starting state $i\geq 0$, the finite-time transition probabilities $P_{i,j}(\tau) = \Pr(X(\tau)=j \mid X(0)=i)$ obey the system of ordinary differential equations
\begin{equation} 
\label{eq:odes}
\begin{split}
\od{P_{i,0}(\tau)}{\tau} &= \mu_1 P_{i,1}(\tau) -\lambda_0 P_{i,0}(\tau) \text{, and} \\
\od{P_{i,j}(\tau)}{\tau} &= \lambda_{j-1}P_{i,j-1}(\tau) + \mu_{j+1}P_{i,j+1}(\tau) - (\lambda_j + \mu_j)P_{i,j}(\tau),
\end{split}
\end{equation}
for $j\geq 1$ with $P_{i,i}(0) = 1$ and $P_{i,j}(0) = 0$ for $i \neq j$ \citep{Feller1971Introduction}.  

For some simple parameterizations of $\lambda_k$ and $\mu_k$, closed-form solutions exist for the transition probabilities $P_{i,j}(\tau)$, but this is not possible for most models.  \citet{Karlin1957Differential} show that for any parameterization of $\lambda_k$ and $\mu_k$, it is possible to express the transition probabilities in terms of orthogonal polynomials.  However, in practice these special polynomials are difficult to find, and even when they are available, they rarely yield solutions in closed-form or expressions that are amenable to computation \citep{Novozhilov2006Biological,Renshaw2011Stochastic}.  In contrast, the continued fraction method we outline below does not require additional model-specific insight beyond specification of $\lambda_k$ and $\mu_k$.

To solve for the transition probabilities, it is advantageous to work in the Laplace domain \citep{Karlin1957Differential}.  This transformation also proves essential in maintaining numerical stability of transition probabilities in general BDPs and in computing the conditional expectations necessary for the EM algorithm derived in a subsequent section.  Laplace transforming equation \eqref{eq:odes} yields
\begin{equation}
\begin{split}
sf_{i,0}(s) - \delta_{i0} &= \mu_1 f_{i,1}(s) - \lambda_0 f_{i,0}(s),\\
sf_{i,j}(s) - \delta_{ij} &= \lambda_{j-1}f_{i,j-1}(s) + \mu_{j+1}f_{i,j+1}(s) - (\lambda_j + \mu_j)f_{i,j}(s),
\end{split}
\label{eq:fijlts}
\end{equation}
where $f_{i,j}(s)$ is the Laplace transform of $P_{i,j}(\tau)$ and $\delta_{ij}=1$ if $i=j$ and zero otherwise.  Letting $i=0$ and rearranging \eqref{eq:fijlts}, we obtain the recurrence relations
\begin{equation}
\begin{split}
f_{0,0}(s) &= \frac{1}{s + \lambda_0 - \mu_1 \left(\frac{f_{0,1}(s)}{f_{0,0}(s)}\right) } \text{, and} \\
\frac{f_{0,j}(s)}{f_{0,j-1}(s)} &= \frac{\lambda_{j-1}}{s + \mu_j + \lambda_j - \mu_{j+1} \left(\frac{f_{0,j+1}(s)}{f_{0,j}(s)}\right)}. 
\end{split}
\label{eq:recur2}
\end{equation}
We can inductively combine these expressions for $j=1,2,3,\ldots$ to arrive at the well-known generalized continued fraction 
\begin{equation}
\begin{array}[t]{r@{\hspace{-0.5em}}l}
\begin{array}{c}
	f_{0,0}(s) = \cfrac{1}{s+\lambda_0 - \cfrac{\lambda_0 \mu_1}{s+\lambda_1+\mu_1 - \cfrac{\lambda_1 \mu_2}{s+\lambda_2+\mu_2 - \cdots}}}
\end{array} 
&
\begin{array}{c}
	\\ \\ \vspace{-1em} .
\end{array}
\end{array}	
\label{eq:cfrac1}	
\end{equation}
This is an exact expression for the Laplace transform of the transition probability $P_{0,0}(\tau)$.  In \eqref{eq:cfrac1}, let $a_1 = 1$ and $a_j = -\lambda_{j-2}\mu_{j-1}$, and let $b_1 = s+\lambda_0$ and $b_j = s+\lambda_{j-1}+\mu_{j-1}$ for $j\geq 2$.  Then \eqref{eq:cfrac1} becomes
\begin{equation}
\begin{array}[t]{r@{\hspace{-0.5em}}l}
\begin{array}{c}
  f_{0,0}(s) = \cfrac{a_1}{b_1 + \cfrac{a_2}{b_2 + \cfrac{a_3}{b_3 + \cdots}}} 
\end{array} 
&
\begin{array}{c}
	\\ \\ \vspace{-1em} .
\end{array}
\end{array}	  
\end{equation}
We can write this more compactly as 
\begin{equation}
 f_{0,0}(s)  = \frac{a_1}{b_1+} \frac{a_2}{b_2+} \frac{a_3}{b_3+} \cdots .
 \label{eq:cfrac2}
\end{equation}
The $k$th convergent of $f_{0,0}(s)$ is
\begin{equation}
 f_{0,0}^{(k)}(s) = \frac{a_1}{b_1+} \frac{a_2}{b_2+} \cdots \frac{a_k}{b_k} = \frac{A_k(s)}{B_k(s)} ,
\label{eq:convergent}
\end{equation}
where $A_k(s)$ and $B_k(s)$ are the numerator and denominator of the rational function $f_{0,0}^{(k)}$.  The transition probabilities $P_{i,j}(\tau)$ for $i,j>0$ can be derived in continued fraction form by combining \eqref{eq:fijlts} and \eqref{eq:cfrac1} to obtain  
\begin{equation}
f_{i,j}(s) = \begin{cases} 
\displaystyle\left(\prod_{k=j+1}^i \mu_k\right)\frac{B_j(s)}{B_{i+1}(s)+} \frac{B_i(s) a_{i+2}}{b_{i+2}+} \frac{a_{i+3}}{b_{i+3}+}\cdots  & \text{for $j\leq i$}, \\
	& \\
\displaystyle\left(\prod_{k=i}^{j-1} \lambda_k\right) \frac{B_i(s)}{B_{j+1}(s)+} \frac{B_j(s) a_{j+2}}{b_{j+2}+} \frac{a_{j+3}}{b_{j+3}+}\cdots  & \text{for $i \leq j$,}
\end{cases}
	\label{eq:fijfull}
\end{equation}
\citep{Murphy1975Some,Crawford2011Transition}.  

Although the Laplace transforms of the transition probabilities are generally still not available in closed-form, a continued fraction representation is desirable for several reasons: 1) continued fraction representations of functions often converge much faster than equivalent power series; 2) there are efficient algorithms for evaluating them to a finite depth; and 3) there exist methods for bounding the error of truncated continued fractions \citep{Bankier1942Numerical,Wall1948Analytic,Blanch1964Numerical,Lorentzen1992Continued,Craviotto1993Survey,Abate1999Computing,Cuyt2008Handbook}.  For an arbitrary BDP, we recover the transition probabilities through numerical inversion of the Laplace-transformed expressions.  We evaluate the continued fraction to a monitored depth that controls the overall error and generates stable approximations to the transition probabilities unattainable by previous methods \citep{Murphy1975Some,Parthasarathy2005Exact,Crawford2011Transition}.

The ability to compute transition probabilities for general BDPs with arbitrary rate parameterizations proves useful in two ways.  First, if we interpret finite-time transition probabilities as functions of an unknown parameter vector $\btheta$, then $P_{a,b}(t)$ given $\btheta$ returns the \emph{likelihood} of a discrete observation from a BDP such that $X(0)=a$ and $X(t)=b$, where the trajectory in time $t$ between $a$ and $b$ is unobserved.  Second, transition probabilities play an important role in computing conditional expectations of sufficient statistics, as we shall see below.

%%%%%%%%%%%%%%%%%%%%%%%%%%%%%%%%%%%%%%%%%%%%%%%%%%%%%%%%

\subsection{Likelihood expressions and surrogate functions}

\begin{figure*}
\centering
\includegraphics{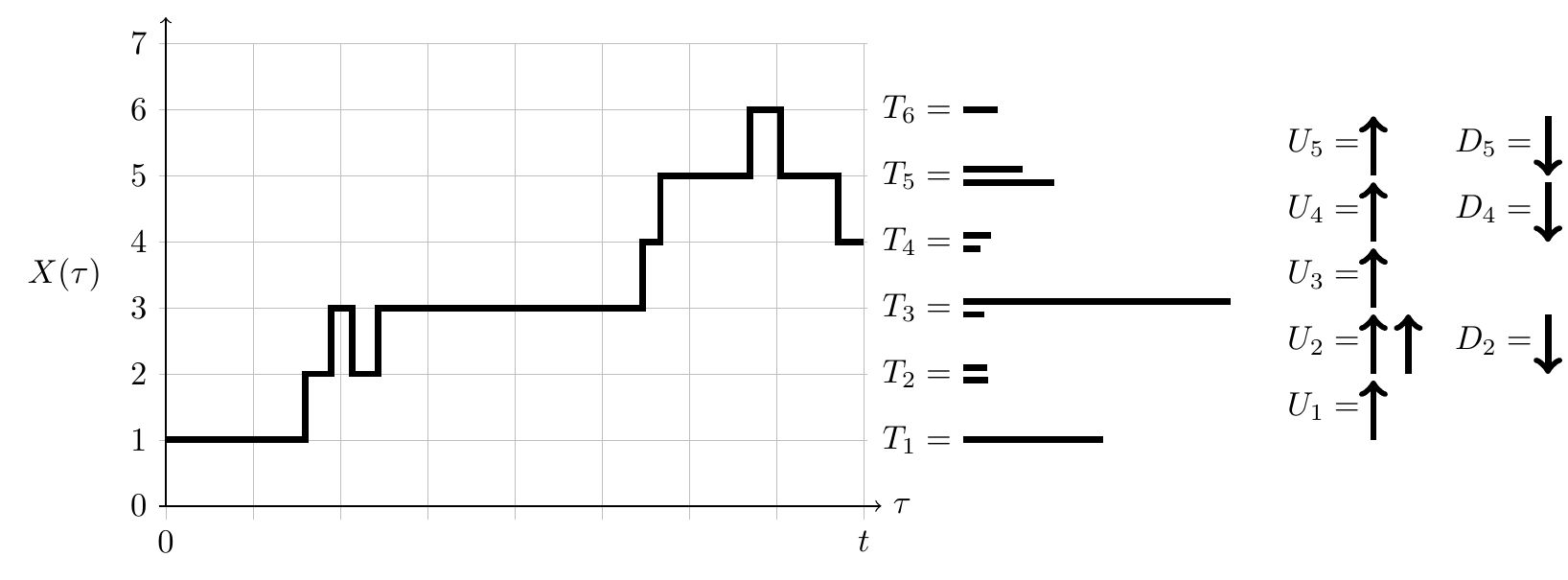}
\caption[A Sample path from a birth-death process]{A sample path from a birth-death process (BDP) $X(\tau)$.  The process starts at state $X(0)=1$ and is at state $X(t)=4$ at time $t$. At right are schematic representations of the time spent in each state $T_k$, the number of up steps $U_k$, and the number of down steps $D_k$.  These quantities are the sufficient statistics for estimators of rate parameters in general birth-death processes.}
\label{fig:bdpath}
\end{figure*}

With a formal description of a general BDP and the finite-time transition probabilities in hand, we now proceed with our task of estimating the parameters of a general BDP using discrete observations.  Given one or more independent observations of the form $\Y=(X(0)=a,X(t)=b)$ from a general BDP, we wish to find maximum likelihood estimates of the rate parameters $\lambda_k$ and $\mu_k$ for $k=0,1,2,\ldots$.  We will assume that the birth and death rates at state $k$ depend on both $k$ and a finite-dimensional parameter vector $\btheta$, so that the form of $\lambda_k(\btheta)$ and $\mu_k(\btheta)$ is known for all $k$.

For a single realization of the process starting at $X(0)=a$ and ending at $X(t)=b$, let $T_k$ be the total time spent in state $k$.  Let $U_k$ be the number of ``up'' steps (births) from state $k$, and let $D_k$ be the number of ``down'' steps (deaths) from state $k$.  Let the total number of up and down steps in a realization of the process be denoted by  $U = \sum_{k=0}^\infty U_k$ and $D = \sum_{k=0}^\infty D_k$ respectively.  We also define the total particle time, 
\begin{equation}
 T_\text{particle} = \int_0^t X(\tau)\dx{\tau} = \sum_{k=0}^\infty k T_k,
\end{equation}
that counts the amount of time lived by each particle since time $\tau=0$.  Of course, the total elapsed time is $t = \sum_{k=0}^\infty T_k$.  We demonstrate these concepts schematically in Figure \ref{fig:bdpath}.

The log-likelihood for a continuously observed process takes a simple form when we sum over all possible states $k$ \citep{Wolff1965Problems}:
\begin{equation}
 %\begin{split}
  \ell(\btheta) = \sum_{k=0}^\infty U_k\log\big[\lambda_k(\btheta)\big] + D_k\log\big[\mu_k(\btheta)\big] - \big[\lambda_k(\btheta) + \mu_k(\btheta)\big] T_k  .
 %\end{split}
	\label{eq:loglik}
\end{equation}
However, when a BDP is sampled discretely such that only $X(0)=a$ and $X(t)=b$ are observed, the quantities $U_k$, $D_k$, and $T_k$ are unknown for every state $k$, and we cannot maximize the log-likelihood \eqref{eq:loglik} without them.  

We therefore appeal to the EM algorithm for iterative maximum likelihood estimation with missing data \citep{Dempster1977Maximum}.  In the EM algorithm, we define a surrogate objective function $Q$ by taking the expectation of the complete data log-likelihood \eqref{eq:loglik}, conditional on the observed data $\Y$ and the parameter values $\btheta^{(m)}$ from the previous iteration of the EM algorithm (the E-step).  Then we find the parameter values $\btheta^{(m+1)}$ that maximize this surrogate function (the M-step).  This two-step process is repeated until convergence to the maximum likelihood estimate of $\btheta$.  Taking the expectation of \eqref{eq:loglik} conditional on $\Y$ and $\btheta^{(m)}$, we form the surrogate function $Q$:
\begin{equation}
	\begin{split}
		Q\big(\btheta\mid\btheta^{(m)}\big) &= \E\big[\ell(\btheta) \mid \Y, \btheta^{(m)}\big] \\
	 &= \sum_{k=0}^\infty \E(U_k|\Y)\log\big[\lambda_k(\btheta)\big] + \E(D_k|\Y)\log\big[\mu_k(\btheta)\big] - \E(T_k|\Y) \big[\lambda_k(\btheta) + \mu_k(\btheta)\big], 
\end{split}
	\label{eq:q}
\end{equation}
where for clarity we have omitted the dependence of the expectations on the parameter value $\btheta^{(m)}$ from the $m$th iterate.  In general, we assume that the maximum likelihood estimator exists; see \citet{Bladt2005Statistical} for a discussion of the issues of identifiability, existence, and uniqueness.

%%%%%%%%%%%%%%%%%%%%%%%%%%%%%%%%%%%%%%%%%%%%%%%%%%%%%%%%%%
\subsection{Computing the expectations of the E-step}

\label{sec:estep}

Computing the expectations of $U_k$, $D_k$, and $T_k$ in the E-step is difficult in birth-death estimation since the unobserved state path and waiting times are not independent conditional on the observed data $\Y$.  \citet{Doss2010Great} adopt an approach for linear BDPs that combines analytic results with simulations.  For some models, these authors are able to derive the generating function for the joint distribution of $U$, $D$, $T_\text{particle}$, and the state path conditional on $X(0)=a$ and can manipulate this generating function to complete the E-step.  For a more complicated linear model, \citeauthor{Doss2010Great} resort to approximating the relevant conditional expectations by simulating sample paths, conditional on $\Y$ \citep{Hobolth2008Markov}.

Our solution is to recognize that we do not need to know very much about the missing data to find the conditional expectations used in the sufficient statistics above.  In fact, the transition probabilities are all that we require.  The following integral representations of the conditional expectations in the EM algorithm will prove useful:
\begin{subequations}
\label{eq:integralexpectations}
	\begin{align}
\E(U_k|\Y) &= \frac{\displaystyle\int_0^t P_{a,k}(\tau)\lambda_k P_{k+1,b}(t-\tau)\dx{\tau}}{P_{a,b}(t)}, \label{eq:euk} \\
 \E(D_k|\Y) &= \frac{\displaystyle\int_0^t P_{a,k}(\tau)\mu_k P_{k-1,b}(t-\tau)\dx{\tau}}{P_{a,b}(t)}, \quad\text{and} \label{eq:edk} \\
 \E(T_k|\Y) &= \frac{\displaystyle\int_0^t P_{a,k}(\tau) P_{k,b}(t-\tau)\dx{\tau}}{P_{a,b}(t)}.\label{eq:etk} 
\end{align}
\end{subequations}
These formulas have appeared in many types of studies related to EM estimation for continuous-time Markov chains \citep{Lange1995Gradient,Holmes2002Expectation,Bladt2005Statistical,Hobolth2005Statistical,Metzner2007Generator}.  For general BDPs whose transition probabilities must be computed numerically, numerical integration over the product of the densities can be computationally prohibitive.

However, the numerators in \eqref{eq:integralexpectations} a-c are convolutions of integrable time-domain functions.  Since the Laplace transforms $f_{a,b}(s)$ of these transition probabilities are available and easy to compute, we take advantage of the Laplace convolution property, arriving at the representations
\begin{subequations}
\label{eq:convolutionexpectations}
	\begin{align}
	\E(U_k|\Y) &= \lambda_k \frac{\mathcal{L}^{-1}\Big[f_{a,k}(s)\ f_{k+1,b}(s) \Big](t)}{P_{a,b}(t)}, \label{eq:eukconv} \\
	\E(D_k|\Y) &= \mu_k \frac{\mathcal{L}^{-1}\Big[f_{a,k}(s)\ f_{k-1,b}(s) \Big](t)}{P_{a,b}(t)},\quad\text{and} \label{eq:edkconv}\\
	\E(T_k|\Y) &= \frac{\mathcal{L}^{-1}\Big[f_{a,k}(s)\ f_{k,b}(s) \Big](t)}{P_{a,b}(t)}.  \label{eq:etkconv}
\end{align}
\end{subequations}
where $\mathcal{L}^{-1}$ denotes inverse Laplace transformation.  Although these formulas are equivalent to \eqref{eq:integralexpectations}, they offer substantial time savings over computing the integral directly, and render tractable the computation of expectations in the EM algorithm for arbitrary general BDPs.  

To calculate the numerators of \eqref{eq:convolutionexpectations}, we use the Laplace inversion method popularized by \citet{Abate1992Numerical,Abate1995Numerical}.  This involves a Riemann sum approximation of the inverse transform that stabilizes the discretization error and is amenable to series acceleration methods \citep{Abate1999Computing,Press2007Numerical}.   To evaluate the continued fraction Laplace transforms $f_{a,b}(s)$, we use the modified Lentz method \citep{Lentz1976Generating,Thompson1986Coulomb,Press2007Numerical}.

%%%%%%%%%%%%%%%%%%%%%%%%%%%%%%%%%%%%%%%%%%%%%%%
\subsection{Maximization techniques for various BDPs}

In contrast to the generic technique outlined above for computing the expectations of the E-step, the M-step depends explicitly on the functional form of the birth and death rates $\lambda_k(\btheta)$ and $\mu_k(\btheta)$.  Here we give several representative examples of BDPs and techniques for completing the M-step of the EM algorithm, such as analytic maximization, minorize-maximize (MM), and Newton's method.  

\subsubsection{Simple linear BDP}

\label{sec:simple}

In the simple linear BDP, births and deaths happen at constant per-capita rates, so $\lambda_k = k\lambda$ and $\mu_k = k\mu$. The unknown parameter vector is $\btheta = (\lambda, \mu)$, and the surrogate function becomes
\begin{equation}
\begin{split}
  Q(\btheta) &= \sum_{k=0}^\infty \E(U_k|\Y)\log[k\lambda] + \E(D_k|\Y)\log[k\mu] - \E(T_k|\Y) k(\lambda+\mu).
\end{split}
  \label{eq:qsimple}
\end{equation}
Taking the derivative of \eqref{eq:qsimple} with respect to the unknown parameters, setting the result to zero, and solving for $\lambda$ and $\mu$ gives the M-step updates 
\begin{subequations}
\label{eq:simpleupdates}
	\begin{align}
		\lambda^{(m+1)} &= \frac{\E(U|\Y)}{\E(T_\text{particle}|\Y) }\text{, and}  \label{eq:simplelambda} \\
    \mu^{(m+1)} &= \frac{\E(D|\Y)}{\E(T_\text{particle}|\Y) }.  \label{eq:simplemu} 
	\end{align}
\end{subequations}
These updates correspond to the usual maximum likelihood estimators in the continuously observed process \citep{Reynolds1973Estimating}.  Note that the transition probabilities $P_{a,b}(t)$ in the denominators of the expectations in \eqref{eq:integralexpectations} cancel out in \eqref{eq:simplelambda} and \eqref{eq:simplemu}.  When this is the case, transition probabilities are not necessary to derive an EM algorithm.

%%%%%%%%%%%%%%%%%%%%%%%%%%%%%%%%%%%%%%%%%%%%%%
\subsubsection{Linear BDP with immigration}

\label{sec:im}

Sometimes populations are not closed, and new individuals can enter; we call this action ``immigration.''  Another interpretation arises in models of point mutations in DNA sequences.  Suppose new mutations arise in a DNA sequence via two distinct processes: one inserts new mutants at a rate proportional to the number already present, and the other creates new mutations at a constant rate, regardless of how many already exist.  To model this behavior, we augment the simple linear BDP above with a constant term $\nu$ representing immigration, so that $\lambda_k = k\lambda+\nu$ and $\mu_k = k\mu$.  The log-likelihood becomes
\begin{equation}
%\begin{split}
   \ell(\btheta) = \sum_{k=0}^\infty U_k\log(k\lambda + \nu) + D_k\log(\mu) - T_k[k(\lambda +\mu) + \nu ].
 %\end{split}
	\label{eq:qim}
\end{equation}
Unfortunately, if we take the derivative of the log-likelihood with respect to $\lambda$ or $\nu$, the unknown appears in the denominator of the terms of the infinite sum.  However, since each summand is a concave function of the unknown parameters, we can separate them in a minorizing function $H$ such that for all $\btheta$, $H\big(\btheta|\btheta^{(m)}\big) \leq \ell(\btheta)$ and $H\big(\btheta^{(m)}|\btheta^{(m)}\big) = \ell\big(\btheta^{(m)}\big)$ as follows: 
\begin{equation}
\begin{split}
  \ell(\btheta) &\geq H\big(\btheta|\btheta^{(m)}\big) \\
        &= \sum_{k=0}^\infty U_k \big[ p_k \log\big(p_k \lambda\big) + (1-p_k)\log\big( (1-p_k) \nu \big)\big] + D_k\log(\mu) - \big[k(\lambda+\mu) + \nu \big]T_k, 
\end{split}
  \label{eq:loglikim}
\end{equation}
where 
\begin{equation}
p_k = \frac{k\lambda^{(m)}}{k\lambda^{(m)} + \nu^{(m)}} .
\label{eq:pk}
\end{equation}
Then letting $Q\big(\btheta\mid \btheta^{(m)}\big) = \E\left(H(\btheta) \mid \Y,\btheta^{(m)}\right)$ be the surrogate function, this minorization forms the basis for an EM algorithm in which a step of the minorize-maximize (MM) algorithm takes the place of the M-step, and the ascent property of the EM algorithm is preserved \citep{Lange2010Numerical}.  Maximizing $Q$ with respect to $\lambda$ and $\nu$ yields the updates
\begin{subequations}
\label{eq:imupdates}
	\begin{align}
		\lambda^{(m+1)} &= \frac{\displaystyle\sum_{k=0}^\infty p_k \E(U_k|\Y)}{\E(T_\text{particle}|\Y)}\text{, and} \label{eq:imlambda} \\
		\nu^{(m+1)} &= \frac{\displaystyle\sum_{k=0}^\infty (1-p_k) \E(U_k|\Y)}{t}.  \label{eq:imnu}
	\end{align}
\end{subequations}
Expression \eqref{eq:imlambda} is similar to \eqref{eq:simplelambda}, the update for $\lambda$ in the simple BDP.  The difference lies in that each $\E(U_k|\Y)$ in this case is weighted by the proportion of additions at state $k$ due to births, not immigrations.  The update for $\mu$ is the same as \eqref{eq:simplemu}.

%%%%%%%%%%%%%%%%%%%%%%%%%%%%%%%%%%%%%%%%%%%%
\subsubsection{Logistic/restricted growth}

\label{sec:logistic}

To illustrate an EM algorithm for more complicated rate specifications in which no MM update is evident and the rates no longer depend on the current state $k$ in a linear way, we examine a model for restricted population growth.  Typical \emph{deterministic} population models often incorporate limitations on population size due to the carrying capacity $K$ of the environment.  One famous example is the logistic model of population growth \citep{Murray2002Mathematical}.  Continuous-time stochastic analogs have previously required a finite cap on population size \citep{Tan1991Stochastic}.  These stochastic models roughly mimic the behavior of the deterministic model for population sizes below $K$, but are limited because they do not allow growth beyond $K$. Here we present a model which supports transient growth beyond the carrying capacity, but where the population size tends to a balance between restricted growth and death.  

Suppose births are cooperative, requiring two parents, but fecundity decays as the number of extant particles increases, and death remains an independent process such that $\lambda_k = \lambda k^2 e^{-\beta k}$ and $\mu_k = k\mu$.  Here, we can interpret the carrying capacity roughly as the population size $k>0$ at which $\lambda_k \approx \mu_k$. Ignoring irrelevant terms, the surrogate function becomes
\begin{equation}
%\begin{split}
  Q\big(\btheta \mid \btheta^{(m)}\big) = \sum_{k=0}^\infty \E(U_k|\Y)[\log(\lambda)-\beta k] + \E(D_k|\Y)\log(\mu) - \E(T_k|\Y)[\lambda k^2 e^{-\beta k} + k\mu] . 
  %\end{split}
  \label{eq:qlogistic}
\end{equation}
Since $\lambda$ and $\beta$ appear together, we opt for a numerical Newton step.  The gradient of $Q$ with respect to these parameters is
\begin{equation}
F = \begin{pmatrix} 
  \displaystyle \frac{\E(U|\Y)}{\lambda} - \sum_{k=0}^\infty k^2 e^{-\beta k}\E(T_k|\Y) \\
 \displaystyle -\sum_{k=0}^\infty \left[k\E(U_k|\Y) + \lambda k^3 e^{-\beta k} \E(T_k|\Y) \right]
\end{pmatrix},
\end{equation}
and the Hessian is
\begin{equation}
H = \begin{pmatrix} 
     \displaystyle -\frac{\E(U|\Y)}{\lambda^2} &  \displaystyle -\sum_{k=0}^\infty k^3 e^{-\beta k}\E(T_k|\Y) \\
       \displaystyle -\sum_{k=0}^\infty k^3 e^{-\beta k}\E(T_k|\Y) & \displaystyle \lambda\sum_{k=0}^\infty k^4 e^{-\beta k} \E(T_k|\Y) .
\end{pmatrix}.
\end{equation}
Then we update these parameters by
\begin{equation}
\begin{pmatrix} \lambda^{(m+1)} \\ \beta^{(m+1)} \end{pmatrix} = 
\begin{pmatrix} \lambda^{(m)} \\ \beta^{(m)} \end{pmatrix} - H^{-1} F .
\end{equation}
The ascent property is preserved when a Newton step is used in place of an exact M-step \citep{Lange1995Gradient}.  The update for $\mu$ is the same as \eqref{eq:simplemu}.

%%%%%%%%%%%%%%%%%%%%%%%%%%%%%%%%%%%%%%%%%%%%%%%%%%%%%%%%

\subsubsection{SIS epidemic models}

\label{sec:sis}

Under a very common epidemic model, members of a finite population of size $N$ are classified as either ``susceptible'' to a given disease or ``infected'' \citep{Bailey1964Elements,Andersson2000Stochastic}.  Susceptibles become infected in proportion to the number of currently infected in the population, and infecteds revert to susceptible status with a certain rate independent of how many infecteds there are.  This idealized susceptible-infectious-susceptible (SIS) infectious disease model specifies a general birth-death process in which we track the number of infecteds.  Let $\lambda_k = \beta k (N-k)/N$ be the rate of new infections when there are already $k$ infected in the population. Let $\mu_k = \gamma k/N$ be the rate of recovery of infecteds to susceptibles.  Then if $\btheta = (\beta,\gamma)$, we have
\begin{equation}
%\begin{split}
 Q\big(\btheta|\btheta^{(m)}\big) = \sum_{k=0}^N \E(U_k|\Y)\log(\beta) + \E(D_k|\Y)\log(\gamma) - \E(T_k|\Y) (k(N-k)\beta + k\gamma)/N,
%\end{split}
\end{equation}
and the update for $\beta$ is
\begin{equation}
\beta^{(m+1)} = \frac{N \E(U|\Y)}{\displaystyle\sum_{k=0}^N (N-k)k\E(T_k|\Y)}.
\end{equation}
The update for $\gamma$ is 
\begin{equation}
\gamma^{(m+1)} = \frac{N\E(D|\Y)}{\displaystyle\E(T_\text{particle}|\Y)}.
\end{equation}

%%%%%%%%%%%%%%%%%%%%%%%%%%%%%%%%%%%%%%%%%%%%%%%%%%%%%%%%%%%%%
\subsubsection{Generalized linear models} 

\label{sec:glm}

Our general framework allows assessment of the influence of covariates on the rates of a general BDP in a novel way.  Suppose we sample observations from independent processes $X_i(\tau)$, $i=1,\ldots,N$ and observe $\Y_i=(X_i(0),X_i(t_i))$ associated with $d$ covariates $\z_i=(z_{i1},\ldots,z_{id})^t$.  These processes may represent different subjects in a study.  We model the birth and death rates $\lambda_{ik}$ and $\mu_{ik}$ for each process/subject $X_i$ as functions of $\z_i$ and unknown $d$-dimensional regression coefficients $\btheta_\lambda$ and $\btheta_\mu$ in a generalized linear model (GLM) framework.  We link
\begin{equation}
\log(\lambda_{ik}) = g(k,\z_i^t \btheta_\lambda) \quad\text{and}\quad \log(\mu_{ik}) = h(k,\z_i^t \btheta_\mu),
\end{equation}
where $g(\cdot)$ and $h(\cdot)$ are scalar-valued functions.  We note the possibility that covariates may differ between $\btheta_\lambda$ and $\btheta_\mu$ through trivial modification; to ease notation, we do not explore this direction.  Given $N$ independent processes, we sum log-likelihoods to arrive at the multiple-subject surrogate function:
\begin{equation}
	\begin{split}
		Q\big(\btheta|\btheta^{(m)}\big) &= \sum_{i=1}^N \sum_{k=0}^\infty \Big[\E(U_k|\Y_i) g(k, \z_i^t \btheta_\lambda) + \E(D_k|\Y_i) h(k, \z_i^t \btheta_\mu) \\
		&\qquad - \E(T_k|\Y_i) \left(e^{g(k,\z_i^t \btheta_\lambda)} + e^{h(k,\z_i^t \btheta_\mu)} \right)\Big].
\end{split}
	\label{eq:Qglm}
\end{equation}
Although we cannot usually maximize this surrogate function for all elements of $(\btheta_\lambda,\btheta_\mu)$ simultaneously, a Newton step is often straightforward to derive.  

As an example, consider generalized linear model extension of the simple linear BDP in which
\begin{equation}
%\begin{split}
\log(\lambda_{ik}) = \log(k) + \z_i^t\btheta_\lambda,\quad \text{and}\quad \log(\mu_{ik}) = \log(k) + \z_i^t\btheta_\mu .
%\end{split}
\end{equation}
Taking the gradient of the corresponding surrogate function $Q$ with respect to the parameters $\btheta_\lambda$ yields
\begin{equation}
	\nabla_{\!\btheta_\lambda} Q = \sum_{i=1}^N \E(U|\Y_i) \z_i - e^{\z_i^t \btheta_\lambda} \E(T_\text{particle}|\Y_i)\z_i 
	\label{eq:gradQ}
\end{equation}
and the second differential (Hessian) of $Q$ is
\begin{equation}
	\mathbf{d}_{\btheta_\lambda}^2 Q = -\sum_{i=1}^N e^{\z_i^t \btheta_\lambda}\E(T_\text{particle}|\Y_i) \z_i \z_i^t .
	\label{eq:d2Q}
\end{equation}
Combining these, we arrive at the Newton step for the parameter vector $\btheta_\lambda$:
\begin{equation}
	\btheta_\lambda^{(m+1)} = \btheta_\lambda^{(m)} - \left( \mathbf{d}_{\btheta_\lambda}^2 Q \right)^{-1}	\nabla_{\!\btheta_\lambda} Q .
	\label{eq:nmupdateglm}
\end{equation}
A similar update can be found for $\btheta_\mu$.  These updates are examples of the gradient EM algorithm for regression in Markov processes described by \citet{Wanek1993Multi} and \citet{Lange1995Gradient}.  It is worth noting that the Hessian matrix $\mathbf{d}_{\btheta_\lambda}^2 Q$ can become ill-conditioned, making it difficult to invert for the Newton step in \eqref{eq:nmupdateglm} for some problems.  Unfortunately there is no quasi-Newton option since in general $\E(T_\text{particle}|\Y)e^{\z_i^t\btheta_\lambda}$ is unbounded.  An alternative to inversion of the Hessian matrix is cyclic coordinate descent in which a Newton step is performed for each coordinate $\btheta_j$ individually. This carries the advantage of avoiding matrix inversion, but convergence is slower and the ascent property must be checked at each Newton step.

%%%%%%%%%%%%%%%%%%%%%%%%%%%%%%%%%%%%%%%%%%%%%

\subsection{Implementation}

Before presenting simulation results and our application to microsatellite evolution, we briefly outline some implementation details that ease our subsequent analyses.

\subsubsection{E-step acceleration}
\label{sec:estepaccel}

The E-step in these EM algorithms for BDP estimation usually involves infinite weighted sums of the conditional expectations $\E(U_k|\Y)$, $\E(D_k|\Y)$, and $\E(T_k|\Y)$.  For example, when estimating $\lambda$ in the simple linear BDP, we must evaluate
\begin{equation}
  \E(U|\Y) = \sum_{k=0}^\infty \E(U_k|\Y) = \frac{\displaystyle\sum_{k=0}^\infty \lambda_k \mathcal{L}^{-1}\Big[f_{a,k}(s)\ f_{k+1,b}(s) \Big](t)}{P_{a,b}(t)}.
\label{eq:eu}
\end{equation}
Fortunately, the conditional expectations of $U_k$, $D_k$, and $T_k$ are usually small for $k \ll \min(a,b)$ and $k \gg \max(a,b)$, so it is possible to replace the infinite sum in \eqref{eq:eu} by a finite one.  We find an additional increase in computational efficiency by exchanging the order of Laplace inversion and summation.  Then \eqref{eq:eu} becomes
\begin{equation}
	\E(U|\Y) \approx \frac{\displaystyle\mathcal{L}^{-1}\left[ \sum_{k=k_\text{min}}^{k_\text{max}} \lambda_k  f_{a,k}(s) f_{k+1,b}(s) \right](t)}{P_{a,b}(t)} ,
	\label{eq:sumformula}
\end{equation}
where we choose $k_\text{min}$ to be the largest $k<\min(a,b)$ such that $\lambda_k|f_{a,k}(s)-f_{k+1,a}|<10^{-8}$ and $k_\text{max}$ to be the first $k>\max(a,b)$ such that $\lambda_k|f_{a,k}(s)f_{k+1,b}(s)|<10^{-8}$.  In practice, we rarely need to compute expectations for $k$ less than $\min(a,b)-10$ or greater than $\max(a,b)+10$.

%%%%%%%%%%%%%%%%%%%%%%%%%%%%%%%%%%%%%%%%%%%%%%%%%%%%%%%%%%%%%%%

\subsubsection{Quasi-Newton acceleration of EM iterates}

\begin{figure*}
\centering

\includegraphics{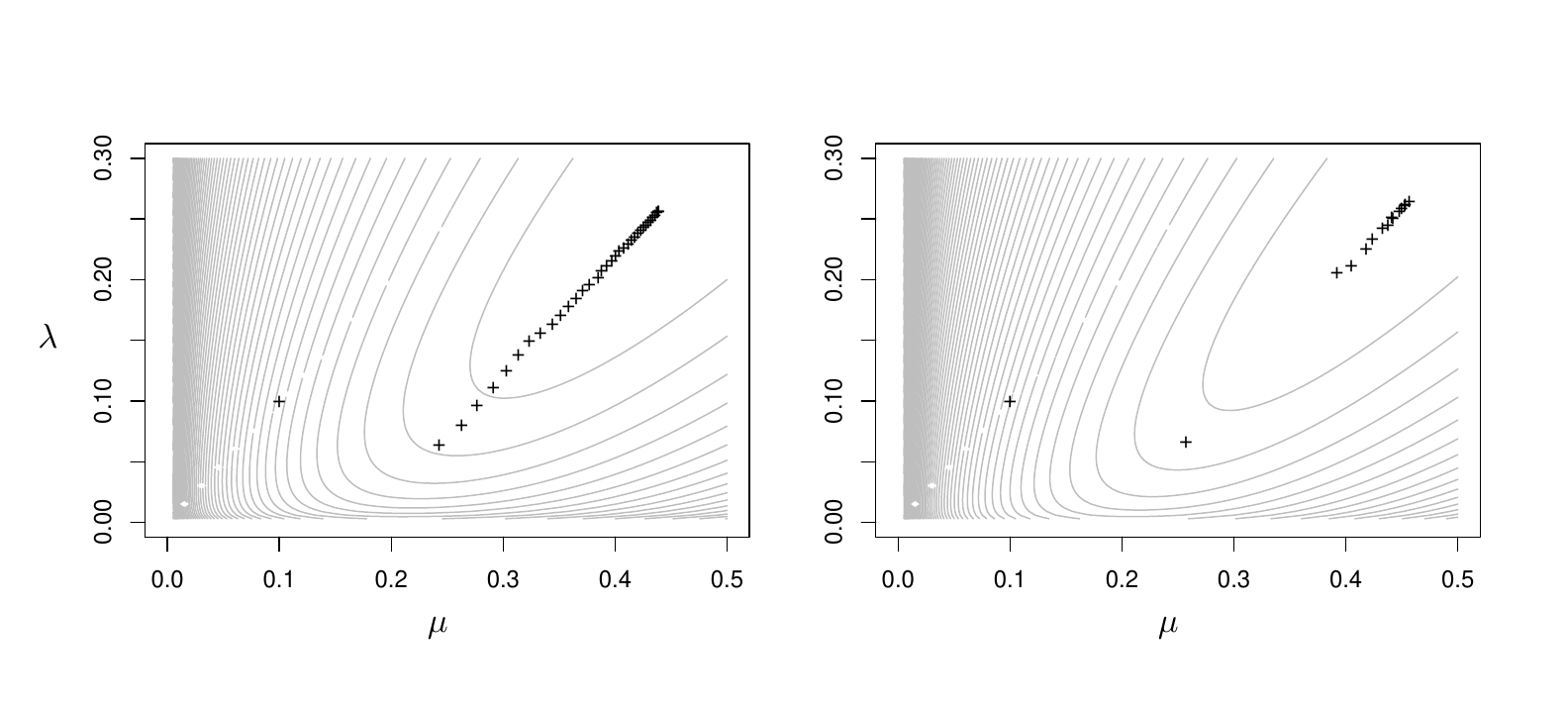}

  \caption[Effect of quasi-Newton acceleration on EM iterates]{Effect of quasi-Newton acceleration on iterates of the expectation-maximization (EM) algorithm for a simple linear BDP with birth rate $\lambda$ and death rate $\mu$.  Contour lines sketch the log-likelihood from $N=50$ discrete samples.  Iterates are shown with the ``$+$'' symbol. On the left, ordinary EM iterates converge very slowly in the neighborhood of the maximum, for a total of 36 iterations.  On the right, EM iterates using quasi-Newton acceleration make large jumps and converge rapidly in 15 iterations.}
\label{fig:emiterates}
\end{figure*}

EM algorithms are notorious for slow convergence, especially near optima.  When appropriate, we exploit the quasi-Newton acceleration method introduced by \citet{Lange1995Quasi} in our implementations.  Other acceleration methods exist, and may give better results, depending on the problem \citep{Lange1995Gradient,Louis1982Finding,Meilijson1989Fast,Jamshidian1993Conjugate}.  Figure \ref{fig:emiterates} shows the log-likelihood function and iterates for the basic EM and accelerated EM methods in the simple linear model.  Since the quasi-Newton acceleration method does not guarantee that the likelihood increases at each step, ``step-halving'' is occasionally necessary to achieve ascent.  Note that this requires likelihood evaluation at least once per iteration.  Our approach is advantageous in that we can efficiently calculate this likelihood (transition probability) for any general BDP \citep{Crawford2011Transition}.

%%%%%%%%%%%%%%%%%%%%%%%%%%%%%%%%%%%%%%%%%%%%%%%%%%%%%%%%%%%%%%

\subsubsection{Asymptotic variance of EM estimates}

Finding the observed information matrix for an EM estimate can be challenging.  \citet{Louis1982Finding} gives formulae for the observed information, which \citet{Doss2010Great} use to derive analytic expressions for the observed information for very simple BDPs. However, analytic expressions for the asymptotic variance are generally hard to find for more complicated models.  We instead turn to the supplemented EM (SEM) algorithm of \citet{Meng1991Using}, which computes the information matrix of the EM estimate of $\btheta$ after the MLE $\hat{\btheta}$ has been found.  The observed information is $\mathbf{I}(\hat{\btheta}) = -\text{d}^2 Q(\hat{\btheta}|\hat{\btheta}) ( \mathbf{I} - \text{d}\mathbf{M}(\hat{\btheta}))$, where $\mathbf{M}(\btheta)$ is the EM algorithm map such that $\btheta^{(m+1)} = \mathbf{M}(\btheta^{(m)})$. We numerically approximate the differential $\text{d}\mathbf{M}$ at the termination of the EM algorithm.  

We note also that since we are able to calculate transition probabilities directly, the observed data log-likelihood is easily computed as
\begin{equation}
 \ell(\btheta) = \sum_{i=1}^N \log P_{a_i,b_i}(t_i), 
\label{eq:obslik}
\end{equation}
where $a_i=X_i(0)$ and $b_i=X_i(t_i)$.  As an alternative to the approaches outlined above, we can calculate the Hessian using purely numerical techniques.  If $\mathbf{H}(\hat{\btheta}) = d^2\ell(\hat{\btheta})$ is the numerical Hessian evaluated at the estimated value $\hat{\btheta}$, then $\hat{\mathbf{I}} \approx -\mathbf{H}(\hat{\btheta})$.

%%%%%%%%%%%%%%%%%%%%%%%%%%%%%%%%%%%%%%%%%%%%%
\section{Results}

\subsection{Laplace convolution E-step comparison}

To illustrate the computational speedup that the Laplace convolution formulae \eqref{eq:convolutionexpectations} and their acceleration in section \ref{sec:estepaccel} achieve over existing methods, we calculate conditional expectations for various BDP models for performing the E-step and report computing times in Table \ref{table:estep}. The first method in the table employs rejection sampling of trajectories where we condition on the starting state, and reject based on the ending state \citep{Bladt2005Statistical}.  The second method adapts an endpoint-conditioned simulation algorithm \citep{Hobolth2008Markov,Hobolth2009Simulation}.  The third considers na\"{i}ve time-domain convolution (Equation \eqref{eq:integralexpectations}) using the \texttt{integrate} function in \texttt{R}.  Finally, we compute the same quantities via the Laplace-domain convolution method outlined in section \ref{sec:estep}.  In our implementations, we have made every effort to reuse as much shared \texttt{R} code as possible, with the aim of making the routines comparable.  We consider four different BDPs.  For a simple linear BDP and a linear BDP with immigration, we use the data $\Y=(X(0)=19,X(2)=27)$.  Under a logistic model, the data are $\Y=(X(0)=10,X(2)=16)$, and for the SIS model the data are $\Y=(X(0)=10,X(2)=31)$.  We list all model parameter values in Table \ref{table:estep}.

As seen in Table \ref{table:estep}, the Laplace convolution method is often more than 10 times faster than the other methods.  In terms of time-performance, the endpoint-conditioned simulation stands as second best, achieving almost comparable speed in the logistic BDP.  To interpret this finding, we recall that \citet{Hobolth2008Markov} constructs an endpoint-conditioned simulation for performing the E-step in finite state-space Markov chains.  Therefore, to adapt this method we approximate each BDP by a Markov chain with a finite transition rate matrix.  To choose the arbitrary dimension of this matrix we truncate the process at the first state $k>\max(a,b)$ such that $P_{a,k}(t)< 10^{-5}$, and the resulting estimates agree substantially with the other methods.  We are aware that the size of the rate matrix  affects the speed of the simulation routine, so we wish to keep the matrix as small as possible.  On the other hand, the matrix must remain large enough to include states that may be visited with high probability in a path from $a$ to $b$ over time $t$.  For the logistic model, such a stringent upper bound lies just above the relatively small carrying capacity.  However, endpoint-conditioned simulation completely fails for the SIS model, an issue we discuss later.  Finally, and quite naturally, the two convolution methods arrived at nearly the same answer for each model; the difference is largely due to very different sources of numerical error, but at disparate computational costs.  

\begin{table*}
%\hspace{-1.5cm}
\centering
\begin{tabular}{llcccc}
\hline
      &          &           & Endpoint-   &             &    \\
      &          & Rejection & conditioned & Time-       & Laplace- \\
Model & Quantity & sampling  & simulation  & convolution & convolution \\
\hline
Simple linear (\ref{sec:simple}) & $\E(U|\Y)$ & 1.449 & 0.741 & 19.606  & 0.084 \\
 $\lambda=0.5$, $\mu=0.3$ & $\E(D|\Y)$        & 1.375 & 0.743 & 21.224 & 0.086 \\
    & $\E(T_\text{particle}|\Y)$              & 1.432 & 0.636 & 16.488 & 0.087 \\
&&&&& \\
Immigration (\ref{sec:im}) & $\sum_kp_k\E(U|\Y)$ & 1.192 & 0.697 & 15.669  & 0.085 \\
$\lambda=0.5$, $\nu=0.2$ & $\E(D|\Y)$         & 1.324 & 0.689 & 21.058 & 0.086 \\
$\mu=0.3$      & $\E(T_\text{particle}|\Y)$   & 1.319 & 0.703 & 14.961  & 0.089 \\
&&&&& \\
Logistic (\ref{sec:logistic}) & $\E(U|\Y)$    & 50.810 & 0.162 & 21.907 & 0.102 \\
$\lambda=0.5$, $\alpha=0.2$ &  $\E(D|\Y)$     & 56.957 & 0.180 & 20.851 & 0.102 \\
$\mu=0.3$ & $\sum_kk^2e^{-k\alpha}\E(T_k|\Y)$ & 50.764 & 0.168 & 21.623  & 0.107 \\
&&&&& \\
SIS (\ref{sec:sis}) & $\E(U|\Y)$              & 7.880 & * & 5.295 & 0.059 \\
$\beta=0.5$, $\gamma=0.3$   & $\E(D|\Y)$      & 8.886 & * & 2.749 & 0.048 \\
            & $\sum_k(N-k)k\E(T_k|\Y)$        & 8.456 & * & 4.269 & 0.053 \\
\hline
\end{tabular}
\caption[Compute times for the E-step in various BDP models]{Compute times (seconds) to perform various E-steps for four different BDP models.  We report text section numbers in which the models are described in parentheses.  For each E-step, we consider several methods.  In all cases, the Laplace method takes substantially less time.  The endpoint-conditioned simulation method fails for the susceptible-infectious-susceptible (SIS) infectious disease model.}
\label{table:estep}
\end{table*}

%%%%%%%%%%%%%%%%%%%%%%%%%%%%%%%%%%%%%%%%%%%%%%%%%%%%

\subsection{Synthetic examples}

To evaluate the performance of our EM algorithms, we simulate discrete observations from several of the BDPs outlined above.  For each sample, we draw starting points $X_i(0)$ uniformly from the integers $0$ to $20$, and times $t_i$ uniformly from $0.1$ to $3$.  We then simulate a trajectory of the BDP and record the state $X_i(t_i)$.  For the generalized linear model (GLM), we employ the simple linear parameterization with a log link with $d=2$ covariates.  We specify the covariates $\z_i=(z_{i,1},z_{i,2})$ as follows:  $z_{i,1} \sim N(1,\sigma^2)$, $z_{i,2} \sim N(2,\sigma^2)$ for $i=1,\ldots,N/2$, $z_{i,1} \sim N(2,\sigma^2)$ and $z_{i,2} \sim N(1,\sigma^2)$ for $i=N/2+1,\ldots,N$, where $\sigma^2=0.1$.

Table \ref{table:sim} reports the number of simulated observations, true parameter values, point-estimates, asymptotic standard error estimates for all model parameters.  It is important to note that the MLEs can differ substantially from the parameter values used to perform the simulation, regardless of the algorithm used to find the MLEs.  This is due to several factors, including: 1) missing state paths; 2) stochasticity of the BDP generating the state paths; 3) arbitrary choice of starting states $X_i(0)$; and 4) finite sample sizes.  Despite these limitations inherent in learning from partially observed stochastic processes, the point-estimates match the true parameter values rather well.

\begin{table}
\centering
\begin{tabular}{lcccc}
\hline
Model                & Parameter & True & Estimate & SE \\
\hline
Simple linear ($N=500$)& $\lambda$   & 0.5 & 0.5039 & 0.0269 \\
 (\ref{sec:simple})  & $\mu$       & 0.2 & 0.1981 & 0.0254 \\
& & & & \\
Immigration ($N=800$)& $\lambda$   & 0.2 & 0.2182 & 0.0129 \\
 (\ref{sec:im})      & $\nu$       & 0.1 & 0.1016 & 0.0213 \\
                     & $\mu$       & 0.25 & 0.2488 & 0.0231 \\
& & & & \\
Logistic ($N=1500$)  & $\lambda$   & 0.3 & 0.2917 & 0.0035 \\
(\ref{sec:logistic}) & $\alpha$    & 0.5 & 0.4942 & 0.0397 \\
                     & $\mu$       & 0.05 & 0.0456 & 0.0633 \\
& & & & \\
SIS ($N=1000$)       & $\beta$     & 0.1 & 0.1025 & 0.0048 \\
 (\ref{sec:sis})     & $\gamma$    & 2.0 & 2.1374 & 0.0367 \\
& & & & \\
GLM ($N=1000$)       & $\btheta_{\lambda,1}$ & 0.25 & 0.2585 & 0.0393 \\
(\ref{sec:glm})      & $\btheta_{\lambda,2}$ & 0.1  & 0.1143 & 0.0402 \\
                     & $\btheta_{\mu,1}$     & 0.2  & 0.1973 & 0.0457 \\
                     & $\btheta_{\mu,2}$     & 0.05 & 0.0877 & 0.0457 \\
\hline
\end{tabular}
\caption[Parameter estimates for various simulated BDPs]{Point-estimates and their standard errors (SE) for simulated observations under various BDPs.  We report the text section describing each of the models in parentheses.  The method for generating the rates in the generalized linear model (GLM) BDP is described in the text.}
\label{table:sim}
\end{table}

%%%%%%%%%%%%%%%%%%%%%%%%%%%%%%%%%%%%%%%%%%%%%%%%%%%%%%%%
\subsection{Application to microsatellite evolution}

Microsatellites are short tandem repeats of characters in a DNA sequence \citep{Schlotterer2000Evolutionary,Ellegren2004Microsatellites,Richard2008Comparative}.  The number of repeated ``motifs'' in a microsatellite often changes over evolutionary timescales.  The molecular mechanism responsible for changes in repeat numbers is known as ``polymerase slippage'' \citep{Schlotterer2000Evolutionary}.  Several researchers have proposed linear BDPs for use in analyzing evolution of microsatellite repeat numbers \citep{Whittaker2003Likelihood,Calabrese2003Dinucleotide,Sainudiin2004Microsatellite}.  However, many investigations demonstrate that microsatellite mutability depends on the number of repeats already present, motif size, and motif nucleotide composition \citep{Chakraborty1997Relative,Eckert2009Every,Kelkar2008Genome,Amos2010Mutation}.  Exactly how these factors affect addition and deletion rates remains an open question \citep{Bhargava2010Mutational}.  

To our knowledge, no previous study formulates or fits a general BDP in which motif size and composition are treated as a covariates in a generalized regression framework, despite the scientific interest in examining such effects on microsatellite evolution.  \citet{Webster2002Microsatellite} study the evolution of 2467 microsatellites common (orthologous) to both humans and chimpanzees, providing an ideal dataset for studying the influence of repeat number and motif size on addition and deletion rates.  For each of these observed microsatellites, \citet{Webster2002Microsatellite} record the motif nucleotide pattern and the number of repeats of this motif found in chimpanzees and humans, and estimate a mutability parameter that controls the rate of addition and deletion.  

We now present an extended application of our BDP inference technique to chimpanzee-human microsatellite evolution, drawing on the data in Table 6 of the supplementary information in \citet{Webster2002Microsatellite}.  We introduce several novel modeling and inferential techniques relevant to the study of microsatellites, and deduce the effect of motif size and composition on microsatellite addition and deletion rates.  While the likelihood takes a slightly more complicated form, our BDP regression technique is straightforward to implement and yields insight into the complicated process of microsatellite evolution.

%%%%%%%%%%%%%%%%%%%%%%%%%%%%%%%%%%%%%%%%%%%%%%%%%%%%%%%%%
\subsubsection{Evolutionary model}

\begin{figure}
\begin{center}

\includegraphics{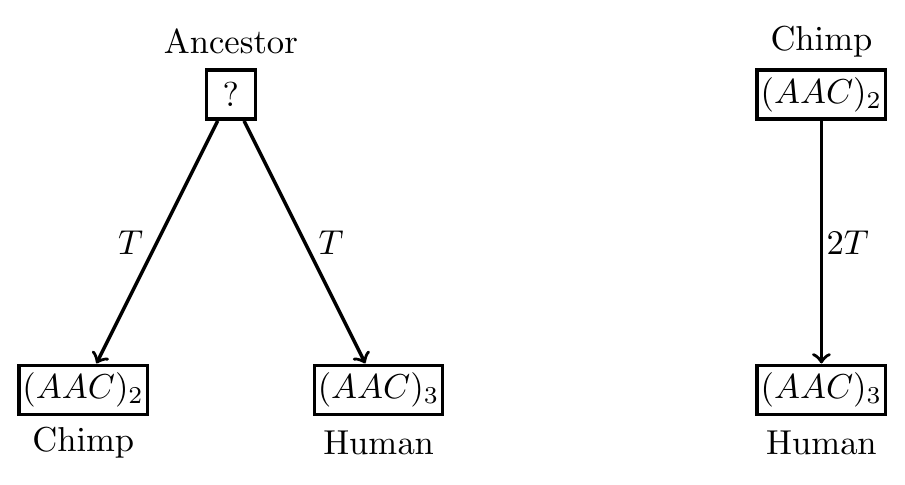}

\end{center}
\caption[Evolutionary relationship of chimpanzees and humans]{Reversibility of the BDP implies that the evolutionary relationship between contemporary chimpanzees and the most recent common ancestor can be inverted.  On the left, the most recent common ancestor of chimpanzees and humans lived at time $T$ in the past.  At a certain locus, chimpanzees have a microsatellite consisting of 2 repeats of the motif $AAC$, and at an orthologous locus, humans have 3 repeats of the motif.  The number of repeats in the ancestor is unknown.  On the right, using a probabilistic justification explained in the text, we may interpret the evolutionary relationship between chimpanzees and humans as unidirectional, while ``integrating out'' the number of repeats at the ancestral locus.}
\label{fig:chimphuman}
\end{figure}

To analyze the data as realizations from a BDP, we must acknowledge the evolutionary relationship between chimpanzees and humans.  Suppose the most recent common ancestor of chimpanzees and humans lived at time $t$ in the past, so that  an evolutionary time of $2t$ separates contemporary humans and chimpanzees.  We note that under mild conditions, general BDPs are reversible Markov chains \citep{Renshaw2011Stochastic}.  Therefore, assuming stationarity of the chimpanzee microsatellite length distributions, we stand justified in reversing the evolutionary process from the ancestor to chimpanzee, so that for estimation purposes we may regard humans as direct descendants of modern chimpanzees (or vice-versa) over an evolutionary time of $2t$.  If $C$ is the number of repeats in a chimpanzee microsatellite and $H$ is the number of repeats in the corresponding human microsatellite, then the likelihood of the observation $\Y=(C,H)$ is
\begin{equation}
\begin{split}
 \Pr(\Y) &= \sum_{k=0}^\infty \pi_k P_{k,C}(t)\ P_{k,H}(t) \\
     &= \pi_{C} \sum_{k=0}^\infty P_{C,k}(t)\ P_{k,H}(t) \\
     &= \pi_{C} P_{C,H}(2t), \\
\end{split}
\label{eq:microsatrev}
\end{equation}
where $\pi_k$ is the equilibrium probability of the microsatellite having $k$ repeats. The second line follows by reversibility and the third by the Chapman-Kolmogorov equality.  Therefore, the log-likelihood of the observation $\Y$ is now $\log \pi_C + \ell(\btheta;\Y)$.  Figure \ref{fig:chimphuman} shows a schematic representation of this reversibility argument.  

%%%%%%%%%%%%%%%%%%%%%%%%%%%%%%%%%%%%%%%%%%%%

\subsubsection{BDP rates and equilibrium distribution}

The observed data for microsatellite $i$ are $\Y_i=(X_i(0), X_i(1))$, where $X_i(0)$ is the number of repeats observed in chimpanzees, $X_i(1)$ is the number of repeats observed in humans, and the evolutionary time separating humans and chimpanzees is scaled to unity.  In addition to the evolutionary relationship explained above, there are other complications: in the \citet{Webster2002Microsatellite} dataset, it is evident that microsatellites with small numbers of repeats are not detected.  \citet{Rose1998Threshold} argue that there is a minimum number of repeats necessary for microsatellite mutation via polymerase slippage.  \citet{Sainudiin2004Microsatellite} interpret this finding as justification for truncating the state-space of BDP at $\xmin$, so that $X(\tau)\geq \xmin$.  To avoid questions of ascertainment bias (see e.g. \citet{Vowles2006Quantifying}), and to make our results comparable to those of past researchers, we \emph{define} a microsatellite to be a collection of more than $x_\text{min}$ repeated motifs, where $\xmin$ is 9 for repeats of size 1, 5 for repeats of size 3 and 4, and 2 for repeats of size 5.  

Researchers have also observed that microsatellites do not tend to grow indefinitely \citep{Kruglyak1998Equilibrium}.  The maximum number of repeats in the \citeauthor{Webster2002Microsatellite} dataset is 47.  This suggests a finite nonzero equilibrium distribution of microsatellite lengths.  To achieve such an equilibrium distribution, we preliminarily view the evolution as a linear BDP with immigration on a state-space that is truncated below $\xmin$.  It is reasonable to assume that rates of addition and deletion depend linearly on how many repeats are already present.  Then for a microsatellite that currently has $k$ repeats, the birth and death rates are
\begin{equation}
\lambda_k = \begin{cases} k\lambda + \lambda & k\geq \xmin \\
                          0                  & k<\xmin \end{cases} \qquad\text{and}\qquad
\mu_k = \begin{cases} k\mu & k>\xmin \\
                      0    & k\leq\xmin . \end{cases} 
\label{eq:microsatrates}
\end{equation}
This gives a geometric equilibrium distribution for the number of repeats:
\begin{equation}
\pi_k = \begin{cases} \displaystyle \left(1-\frac{\lambda}{\mu}\right) \left(\frac{\lambda}{\mu}\right)^{k-\xmin-1}  & k\geq\xmin \\ 0 & k<\xmin , \end{cases}
\label{eq:microsatequilib}
\end{equation}
when $\lambda<\mu$ \citep{Renshaw2011Stochastic}.  We choose this simple model so that the BDP has a simple closed-form nonzero equilibrium solution that is easy to incorporate into the log-likelihood.  Note that the constraint $\lambda<\mu$ does not mean that the rate of microsatellite repeat addition is always less than the rate of deletion, since it is possible that $\lambda_k > \mu_k$ for small $k$.  Additionally, $\lambda<\mu$ does not mean that the number of repeats in a microsatellite tends to zero over long evolutionary times --- the equilibrium distribution \eqref{eq:microsatequilib} assigns positive probability to all repeat numbers greater than or equal to $\xmin$.

%%%%%%%%%%%%%%%%%%%%%%%%%%%%%%%%%%%%%%%%%%%%%%%%%%%%%%%%%%

\subsubsection{Likelihood and surrogate function}

Now we augment the log-likelihood with the log-equilibrium probability of observing $X_i(0)$ chimpanzee repeats
\begin{equation}
F(\btheta) = \sum_{i=1}^N \log \pi_{X_i(0)} + \ell(\btheta; \Y_i) ,
\label{eq:microsatlik}
\end{equation}
where $\ell(\btheta;\Y_i)$ is equivalent to \eqref{eq:loglik}.  Including the influence of the equilibrium distribution is similar to imposing a prior distribution on $\lambda$ and $\mu$. To ensure the existence of the equilibrium distribution \eqref{eq:microsatequilib}, we must also incorporate the constraint $\lambda<\mu$.  To achieve maximization of the augmented log-likelihood \eqref{eq:microsatlik} under this constraint, we impose a barrier term of the form $\gamma \log(\mu-\lambda)$.  By iteratively maximizing and sending the barrier penalty $\gamma\to 0$, we can achieve maximization under the inequality constraint.  More formally, if we let
\begin{equation}
H(\btheta) = \sum_{i=1}^N \left[ \log \pi_{X_i(0)} + \ell(\btheta; \Y_i )\right] + \gamma\log(\mu-\lambda) ,
\end{equation}
then
\begin{equation}
\argmax{\btheta} H(\btheta) \to \argmax{\btheta} F(\btheta)
\end{equation} 
under the constraint $\lambda<\mu$ as $\gamma\to 0$.

To incorporate and evaluate the influence of motif size and composition heterogeneity, we now treat $\lambda$ and $\mu$ in the $i$th observation as functions of the covariate vector $\z_i$ in a general BDP.  Suppose microsatellite $i$ has motif size $r_i$.  We code the vectors $\z_i$ as follows: 
\begin{equation} 
 \z_i=\begin{cases} 
(1,0,0,p_a,p_c,p_t)^t & r_i=1 \\
(1,1,0,p_a,p_c,p_t)^t & r_i=2 \\
(1,0,1,p_a,p_c,p_t)^t & r_i\geq3 \end{cases} \\
\end{equation}
where $p_x$ is the proportion of $x$ nucleotides per repeat.  We define a single parameter $\alpha$ that controls the difference between $\lambda$ and $\mu$.  Then in the $i$th microsatellite, the complete model becomes
\begin{equation}
\log(\lambda_{k,i}) = \log(k+1) + \alpha + \z_i^t\btheta \quad\text{and}\quad \log(\mu_{k,i}) = \log(k) + \z_i^t\btheta.
\end{equation}
Therefore $(\alpha,\btheta)^t$ is the $7\times 1$ vector of unknown parameters.  Putting all this together, the surrogate function becomes
\begin{equation}
\begin{split}
  Q\big(\btheta|\btheta^{(m)}\big) &\propto \Bigg( \sum_{i=1}^N X_i(0)\alpha + \log\left(1 - e^\alpha \right) + \Bigg[ \sum_{k=0}^\infty \E(U_k|\Y_i)(\alpha + \z_i^t\btheta) + \E(D_k|\Y_i)\z_i^t \btheta \\
  & \quad\quad- \E(T_k|\Y_i) \left( (k+1)e^{\alpha + \z_i^t \btheta} + ke^{\z_i^t \btheta} \right)\Bigg]\Bigg)  + \gamma \log(-\alpha) ,
\end{split}
 \label{eq:Qmicrosatglm}
\end{equation}
where $\alpha<0$ since $\lambda<\mu$, and we send the penalty $\gamma\to 0$ as the algorithm converges.  We use a gradient EM algorithm to find the MLE of $(\alpha,\btheta)$.  

Table \ref{table:microsat} reports the parameter estimates, along with asymptotic standard errors.   From these results, we infer that motifs of different sizes and composition have different characteristics under our evolutionary model.  Specifically, $\lambda$ and $\mu$ are greatest for dinucleotide repeats, as compared to motifs with one or at least three repeats.  Motifs consisting mostly of $A$ and $T$ nucleotides also give rise to higher $\lambda$ and $\mu$.  Table \ref{table:unique} shows the estimated $\lambda$ and $\mu$ for each unique motif pattern in the dataset.  These conclusions are largely consistent with the descriptive results obtained by \citet{Webster2002Microsatellite}. Our analysis also provides a natural probabilistic justification for the existence of a finite nonzero equilibrium distribution of microsatellite repeat numbers and a formal statistical framework for deducing the effect of motif size and repeat number on mutation rates.

\begin{table}
\centering
\begin{tabular}{cccc}
\hline
Parameter   & Covariate     & Estimate & SE     \\
\hline
$\btheta_1$  & Intercept    & -1.3105 & 0.1236 \\
$\btheta_2$  & $r_i=2$      &  0.2854 & 0.0983 \\
$\btheta_3$  & $r_i\geq3$   & -1.5405 & 0.1079 \\
$\btheta_4$  & $p_a$        &  0.2207 & 0.1725 \\
$\btheta_5$  & $p_c$        & -0.3822 & 0.0577 \\
$\btheta_6$  & $p_t$        &  0.0477 & 0.0002 \\
$\alpha$     & birth        & -0.0889 & 0.0039 \\
\hline
\end{tabular}
\caption[Maximum likelihood estimates of parameters for the microsatellite model]{Maximum likelihood estimates of parameters in the microsatellite model and their asymptotic standard errors.  The first three elements of $\btheta$ correspond to the motif size $r_i$, and the last three correspond to the motif nucleotide composition.  The parameter $\alpha$ controls the difference between the birth and death rates.  The $i$th microsatellite birth rate is then $\lambda=\exp(\alpha+\z_i^t\btheta)$ and the death rate is $\mu=\exp(z_i^t\btheta)$.  Estimated birth and death rates are higher for dinucleotide repeats than for mononucleotide repeats or microsatellites whose motifs have 3, 4, or 5 nucleotides. Mircrosatellites whose motif consists, for example, of $A$ nucleotides have higher birth and death rates compared to $G$ nucleotides.}
\label{table:microsat}
\end{table}

%\begin{sideways}
\begin{sidewaystable}
\centering
\tiny
\begin{tabular}{lcc|lcc|lcc|lcc|lcc}
\hline 
Motif & $\lambda$ & $\mu$ & Motif & $\lambda$ & $\mu$ & Motif & $\lambda$ & $\mu$ & Motif & $\lambda$ & $\mu$ & Motif & $\lambda$ & $\mu$ \\ 
\hline 
$A$ & 0.3605 & 0.3969 & $AGGA$ & 0.025 & 0.0276 & $CCAT$ & 0.0051 & 0.0056 & $GGCG$ & 0.0133 & 0.0147 & $TCTT$ & 0.0085 & 0.0094 \\ 
$AAAAC$ & 0.0128 & 0.0141 & $AGGG$ & 0.0266 & 0.0293 & $CCATC$ & 0.004 & 0.0044 & $GGCGG$ & 0.0155 & 0.0171 & $TCTTT$ & 0.0096 & 0.0106 \\ 
$AAAAG$ & 0.0233 & 0.0257 & $AGGGA$ & 0.0256 & 0.0282 & $CCT$ & 0.0031 & 0.0035 & $GGGA$ & 0.0266 & 0.0293 & $TG$ & 0.6094 & 0.6708 \\ 
$AAAAT$ & 0.0207 & 0.0228 & $AGTC$ & 0.0108 & 0.0119 & $CCTC$ & 0.0026 & 0.0028 & $GGGAA$ & 0.0256 & 0.0282 & $TGA$ & 0.0214 & 0.0235 \\ 
$AAAC$ & 0.0111 & 0.0123 & $AGTG$ & 0.0229 & 0.0252 & $CCTCC$ & 0.0023 & 0.0025 & $GGGGA$ & 0.0269 & 0.0296 & $TGAA$ & 0.0216 & 0.0237 \\ 
$AAACA$ & 0.0128 & 0.0141 & $AT$ & 0.5407 & 0.5952 & $CCTG$ & 0.0054 & 0.006 & $GGT$ & 0.0231 & 0.0255 & $TGAGT$ & 0.0212 & 0.0233 \\ 
$AAACC$ & 0.0074 & 0.0081 & $ATA$ & 0.0197 & 0.0217 & $CCTT$ & 0.0047 & 0.0052 & $GGTA$ & 0.0229 & 0.0252 & $TGAT$ & 0.0197 & 0.0217 \\ 
$AAAG$ & 0.0236 & 0.026 & $ATAA$ & 0.0203 & 0.0224 & $CCTTT$ & 0.006 & 0.0066 & $GGTG$ & 0.0243 & 0.0268 & $TGC$ & 0.0085 & 0.0094 \\ 
$AAAGA$ & 0.0233 & 0.0257 & $ATAAA$ & 0.0207 & 0.0228 & $CG$ & 0.1835 & 0.202 & $GGTGT$ & 0.0222 & 0.0245 & $TGCC$ & 0.0054 & 0.006 \\ 
$AAAGG$ & 0.0244 & 0.0269 & $ATAC$ & 0.0102 & 0.0112 & $CGC$ & 0.0038 & 0.0042 & $GT$ & 0.6094 & 0.6708 & $TGG$ & 0.0231 & 0.0255 \\ 
$AAAT$ & 0.0203 & 0.0224 & $ATAG$ & 0.0216 & 0.0237 & $CGG$ & 0.0104 & 0.0114 & $GTACA$ & 0.0125 & 0.0138 & $TGGA$ & 0.0229 & 0.0252 \\ 
$AAATA$ & 0.0207 & 0.0228 & $ATATG$ & 0.0202 & 0.0222 & $CT$ & 0.1362 & 0.1499 & $GTAT$ & 0.0197 & 0.0217 & $TGGG$ & 0.0243 & 0.0268 \\ 
$AAATG$ & 0.0217 & 0.0239 & $ATC$ & 0.0079 & 0.0087 & $CTC$ & 0.0031 & 0.0035 & $GTG$ & 0.0231 & 0.0255 & $TGT$ & 0.019 & 0.0209 \\ 
$AAATT$ & 0.0193 & 0.0212 & $ATCT$ & 0.0093 & 0.0103 & $CTCC$ & 0.0026 & 0.0028 & $GTGAG$ & 0.0239 & 0.0263 & $TGTA$ & 0.0197 & 0.0217 \\ 
$AAC$ & 0.0089 & 0.0098 & $ATG$ & 0.0214 & 0.0235 & $CTCCT$ & 0.0037 & 0.0041 & $GTGG$ & 0.0243 & 0.0268 & $TGTC$ & 0.0099 & 0.0109 \\ 
$AACA$ & 0.0111 & 0.0123 & $ATGA$ & 0.0216 & 0.0237 & $CTG$ & 0.0085 & 0.0094 & $GTT$ & 0.019 & 0.0209 & $TGTT$ & 0.018 & 0.0199 \\ 
$AACAA$ & 0.0128 & 0.0141 & $ATGAC$ & 0.0125 & 0.0138 & $CTGGG$ & 0.0138 & 0.0151 & $GTTG$ & 0.0209 & 0.0231 & $TGTTT$ & 0.0175 & 0.0193 \\ 
$AACC$ & 0.0056 & 0.0062 & $ATGAT$ & 0.0202 & 0.0222 & $CTT$ & 0.007 & 0.0077 & $GTTT$ & 0.018 & 0.0199 & $TTA$ & 0.0175 & 0.0193 \\ 
$AACT$ & 0.0102 & 0.0112 & $ATT$ & 0.0175 & 0.0193 & $CTTC$ & 0.0047 & 0.0052 & $GTTTA$ & 0.0188 & 0.0207 & $TTAA$ & 0.0186 & 0.0205 \\ 
$AAG$ & 0.0241 & 0.0265 & $ATTA$ & 0.0186 & 0.0205 & $CTTT$ & 0.0085 & 0.0094 & $GTTTG$ & 0.0197 & 0.0217 & $TTAAT$ & 0.0179 & 0.0197 \\ 
$AAGA$ & 0.0236 & 0.026 & $ATTC$ & 0.0093 & 0.0103 & $CTTTC$ & 0.006 & 0.0066 & $GTTTT$ & 0.0175 & 0.0193 & $TTAG$ & 0.0197 & 0.0217 \\ 
$AAGC$ & 0.0118 & 0.013 & $ATTG$ & 0.0197 & 0.0217 & $CTTTT$ & 0.0096 & 0.0106 & $T$ & 0.2524 & 0.2778 & $TTAT$ & 0.017 & 0.0187 \\ 
$AAGG$ & 0.025 & 0.0276 & $ATTT$ & 0.017 & 0.0187 & $G$ & 0.4579 & 0.5041 & $TA$ & 0.5407 & 0.5952 & $TTATT$ & 0.0167 & 0.0184 \\ 
$AAGGG$ & 0.0256 & 0.0282 & $ATTTA$ & 0.0179 & 0.0197 & $GA$ & 0.7283 & 0.8017 & $TAA$ & 0.0197 & 0.0217 & $TTC$ & 0.007 & 0.0077 \\ 
$AAGT$ & 0.0216 & 0.0237 & $ATTTC$ & 0.0103 & 0.0114 & $GAA$ & 0.0241 & 0.0265 & $TAAA$ & 0.0203 & 0.0224 & $TTCA$ & 0.0093 & 0.0103 \\ 
$AAGTG$ & 0.0228 & 0.0251 & $ATTTG$ & 0.0188 & 0.0207 & $GAAA$ & 0.0236 & 0.026 & $TAAAA$ & 0.0207 & 0.0228 & $TTCC$ & 0.0047 & 0.0052 \\ 
$AAT$ & 0.0197 & 0.0217 & $ATTTT$ & 0.0167 & 0.0184 & $GAAAA$ & 0.0233 & 0.0257 & $TAAAT$ & 0.0193 & 0.0212 & $TTCT$ & 0.0085 & 0.0094 \\ 
$AATA$ & 0.0203 & 0.0224 & $C$ & 0.0229 & 0.0252 & $GAAAG$ & 0.0244 & 0.0269 & $TAAT$ & 0.0186 & 0.0205 & $TTCTC$ & 0.006 & 0.0066 \\ 
$AATAA$ & 0.0207 & 0.0228 & $CA$ & 0.1628 & 0.1792 & $GAAG$ & 0.025 & 0.0276 & $TAATG$ & 0.0202 & 0.0222 & $TTCTG$ & 0.0108 & 0.0119 \\ 
$AATAG$ & 0.0217 & 0.0239 & $CAA$ & 0.0089 & 0.0098 & $GAAGG$ & 0.0256 & 0.0282 & $TAATT$ & 0.0179 & 0.0197 & $TTCTT$ & 0.0096 & 0.0106 \\ 
$AATG$ & 0.0216 & 0.0237 & $CAAA$ & 0.0111 & 0.0123 & $GAAT$ & 0.0216 & 0.0237 & $TAC$ & 0.0079 & 0.0087 & $TTG$ & 0.019 & 0.0209 \\ 
$AATT$ & 0.0186 & 0.0205 & $CAAAA$ & 0.0128 & 0.0141 & $GAATT$ & 0.0202 & 0.0222 & $TACTA$ & 0.0111 & 0.0122 & $TTGAA$ & 0.0202 & 0.0222 \\ 
$AC$ & 0.1628 & 0.1792 & $CAAAC$ & 0.0074 & 0.0081 & $GACAG$ & 0.0141 & 0.0155 & $TAGA$ & 0.0216 & 0.0237 & $TTGT$ & 0.018 & 0.0199 \\ 
$ACA$ & 0.0089 & 0.0098 & $CAAAG$ & 0.0134 & 0.0148 & $GAG$ & 0.0261 & 0.0287 & $TAT$ & 0.0175 & 0.0193 & $TTGTT$ & 0.0175 & 0.0193 \\ 
$ACAA$ & 0.0111 & 0.0123 & $CAC$ & 0.0035 & 0.0039 & $GAGAA$ & 0.0244 & 0.0269 & $TATC$ & 0.0093 & 0.0103 & $TTTA$ & 0.017 & 0.0187 \\ 
$ACAAA$ & 0.0128 & 0.0141 & $CACAC$ & 0.0042 & 0.0047 & $GAGG$ & 0.0266 & 0.0293 & $TATG$ & 0.0197 & 0.0217 & $TTTAA$ & 0.0179 & 0.0197 \\ 
$ACAG$ & 0.0118 & 0.013 & $CACC$ & 0.0028 & 0.0031 & $GAT$ & 0.0214 & 0.0235 & $TATT$ & 0.017 & 0.0187 & $TTTAG$ & 0.0188 & 0.0207 \\ 
$ACC$ & 0.0035 & 0.0039 & $CACCA$ & 0.0042 & 0.0047 & $GATA$ & 0.0216 & 0.0237 & $TATTT$ & 0.0167 & 0.0184 & $TTTAT$ & 0.0167 & 0.0184 \\ 
$ACCA$ & 0.0056 & 0.0062 & $CAG$ & 0.0096 & 0.0106 & $GATT$ & 0.0197 & 0.0217 & $TC$ & 0.1362 & 0.1499 & $TTTC$ & 0.0085 & 0.0094 \\ 
$AG$ & 0.7283 & 0.8017 & $CAGA$ & 0.0118 & 0.013 & $GCACA$ & 0.0077 & 0.0085 & $TCAAA$ & 0.0119 & 0.0131 & $TTTCC$ & 0.006 & 0.0066 \\ 
$AGAA$ & 0.0236 & 0.026 & $CAGAG$ & 0.0141 & 0.0155 & $GCCGC$ & 0.0047 & 0.0051 & $TCAC$ & 0.0051 & 0.0056 & $TTTCT$ & 0.0096 & 0.0106 \\ 
$AGAAA$ & 0.0233 & 0.0257 & $CAGG$ & 0.0126 & 0.0138 & $GCT$ & 0.0085 & 0.0094 & $TCAT$ & 0.0093 & 0.0103 & $TTTG$ & 0.018 & 0.0199 \\ 
$AGAC$ & 0.0118 & 0.013 & $CAT$ & 0.0079 & 0.0087 & $GCTGT$ & 0.0122 & 0.0134 & $TCATT$ & 0.0103 & 0.0114 & $TTTGG$ & 0.0197 & 0.0217 \\ 
$AGAGG$ & 0.0256 & 0.0282 & $CATA$ & 0.0102 & 0.0112 & $GGA$ & 0.0261 & 0.0287 & $TCC$ & 0.0031 & 0.0035 & $TTTGT$ & 0.0175 & 0.0193 \\ 
$AGAT$ & 0.0216 & 0.0237 & $CATC$ & 0.0051 & 0.0056 & $GGAA$ & 0.025 & 0.0276 & $TCCA$ & 0.0051 & 0.0056 & $TTTTA$ & 0.0167 & 0.0184 \\ 
$AGCAA$ & 0.0134 & 0.0148 & $CATG$ & 0.0108 & 0.0119 & $GGAAG$ & 0.0256 & 0.0282 & $TCCC$ & 0.0026 & 0.0028 & $TTTTC$ & 0.0096 & 0.0106 \\ 
$AGCC$ & 0.0059 & 0.0065 & $CATT$ & 0.0093 & 0.0103 & $GGAG$ & 0.0266 & 0.0293 & $TCCT$ & 0.0047 & 0.0052 & $TTTTG$ & 0.0175 & 0.0193 \\ 
$AGCTC$ & 0.0072 & 0.0079 & $CCA$ & 0.0035 & 0.0039 & $GGAGG$ & 0.0269 & 0.0296 & $TCT$ & 0.007 & 0.0077 &  \\ 
$AGG$ & 0.0261 & 0.0287 & $CCAAC$ & 0.0042 & 0.0047 & $GGCCA$ & 0.0081 & 0.0089 & $TCTG$ & 0.0099 & 0.0109 &  \\ 
\hline 
\end{tabular}
\caption{Estimates of birth and death rates for each unique motif in the human-chimpanzee dataset of \citet{Webster2002Microsatellite}. Under our model of microsatellite mutation, $AT$ repeats have the highest associated rates and $CTCC$ repeats have the lowest. }
\label{table:unique}
\end{sidewaystable}
%\end{sideways}

%%%%%%%%%%%%%%%%%%%%%%%%%%%%%%%%%%%%%%%%%%%%%%%%%%%%%%%%%
\section{Discussion}

Application of stochastic models in statistics requires a flexible and general approach to parameter estimation, without which even the most realistic model becomes unappealing to researchers who wish to learn from the data they have collected.  Estimation for continuously observed BDPs is straightforward and well-established.  For partially observed BDPs, our approach is unique because it requires only two simple ingredients: the functional form of the birth and death rates $\lambda_k(\btheta)$ and $\mu_k(\btheta)$ for all $k$, and an exact or approximate M-step.  A third ingredient is optional: the Hessian of the surrogate function is useful when asymptotic standard errors are desired.  However, this matrix can often be approximated numerically upon convergence of the EM algorithm, since the observed-data likelihood is available numerically via \eqref{eq:obslik}.  With these ingredients in hand, even elusive general BDPs become tractable.

In previous work on estimation for BDPs, completion of the E-step typically relies on time-domain numerical integration or simulation of BDP trajectories.  As we show in Table \ref{table:estep}, both rejection sampling and endpoint-conditioned simulation can occasionally perform satisfactorily, especially in comparison to time-domain convolution.  However, endpoint-conditioning is designed for finite state-space Markov chains, and it relies on a matrix eigendecomposition to calculate transition probabilities.  As we show for the SIS model, this matrix becomes nearly singular, causing the simulation algorithm to fail, even when we choose parameter values that are not biologically unreasonable.  The Laplace convolution in the E-step of our algorithm is more generic with equivalent or better performance.  For this reason, a variation on our Laplace convolution method for computing the E-step may offer further use in estimation for non-BDP finite Markov chains as well, such as nucleotide or codon substitution models.  For some linear BDPs, the availability of a generating function furnishes analytic E- and M-steps yielding very fast parameter updates in closed-form \citep{Doss2010Great}.  For some models, these tools provide the asymptotic variance of the MLE in closed-form.  However, for the majority of BDPs, we must return to the Laplace convolution method outlined in this paper.

If one cannot find analytic parameter updates in the M-step, several options remain available.  With a minorizing function as in section \ref{sec:im}, an EM-MM algorithm is viable.  Further, one or more numerical Newton steps offers an alternative, as in sections \ref{sec:logistic} and \ref{sec:glm}.  One may employ other gradient-based methods as well.  Although the MM update derived for the BDP with immigration (section \ref{sec:im}) is appealing in its simplicity, multiple minorizations of the likelihood can result in very slow convergence, since the surrogate function lies far from the true likelihood for most values of $\btheta$.  In addition, Newton steps that require matrix inversion may suffer since the Hessian of the surrogate can become ill-conditioned.  

Even with the substantial speedup offered by our Laplace convolution method for performing the E-step and quasi-Newton acceleration of the EM iterates, our algorithms can move slowly toward the MLE.  Here, na\"{i}ve numerical optimization of the incomplete data likelihood can sometimes run computationally faster. However, such techniques perform very poorly when the number of parameters increases and they often require specification of tuning constants in order to reach the global optimum.  For BDP estimation problems, EM algorithms offer several other advantages over na\"{i}ve numerical optimization, and these benefits are especially stark when the M-step is available in closed-form.  First, when the log-likelihood is locally convex, the EM algorithm is robust with respect to the initial parameter values near the maximum, and EM algorithms generally do not need tuning parameters.  Further, the ascent property ensures the iterates will approach a maximum.  Perhaps the most important reason to consider EM algorithms is that they can accommodate high-dimensional parameter spaces without substantially increasing the computational complexity of the algorithm.  This is especially useful in models with many unknown parameters when performing regression with covariates (section \ref{sec:glm}), or our microsatellite example.  We also note the potential for substantial computational speedup by parallelizing the E-step.  When discrete observations from a BDP are independent, the E-step may be performed in parallel for every observation.  For example, $\E(U|\Y_i)$ can be computed simultaneously for $i=1,\ldots,N$.  When speed is an issue, graphics processing units may prove useful in reducing the computational cost of EM algorithms \citep{Zhou2010Graphics}.

With regard to our example, we present a novel way of studying the evolution of microsatellite repeats using a generalized linear model.  Previous efforts often ignore the evolutionary relationship between organisms, use incomplete or equilibrium models of repeat numbers, or fit separate models to motifs of different sizes.  We treat motif size as a categorical variable and incorporate motif nucleotide composition, allowing us to fit a single model to all the microsatellite observations simultaneously.  Though our rate specification \eqref{eq:microsatrates} and resulting equilibrium distribution \eqref{eq:microsatequilib} are intended to be somewhat simplistic, more sophisticated models that are informed by biological considerations may be fruitful.  The only requirement in our setup is that the gradient and Hessian of $\lambda_k$, $\mu_k$, and $\pi_k$ be available for any repeat number $k$.  Although our microsatellite example is limited in scope, it is easy to imagine a more comprehensive study.  For example, incorporating more sophisticated motif nucleotide composition covariates and location of the microsatellite on the chromosome might provide additional insight into the evolutionary process.  Our EM framework is nearly ideal for these types of studies, since the number of unknown parameters does not substantially increase the computational burden of the M-step, and the E-step is completely unaffected by the number of parameters.  

Interestingly, we attempted to use the generic nonlinear regression \texttt{R} function \texttt{nlm} to validate the MLEs obtained by our EM algorithm for the microsatellite evolution problem, starting at a variety of initial values, including the MLE found by our EM algorithm.  This na\"{i}ve optimizer failed to converge in every case.  We speculate that this is because the small numerical errors in the likelihood evaluation have similar order of magnitude as the curvature of the likelihood function near the maximum.  Our EM aglorithms take advantage of analytic derivatives of the surrogate function instead of the likelihood, and hence are less susceptible to small errors in the numerical gradient.

%%%%%%%%%%%%%%%%%%%%%%%%%%%%%%%%%%%%%%%%%%%%%%%%%%%%%%%%%
\section{Conclusion}

Previous work on parameter estimation in BDPs almost exclusively confines itself to inference of birth and death rates under the simple linear model.  To rectify this situation, we present a flexible and robust framework for deriving EM algorithms to estimate parameters in any general BDP, using discrete observations.  We hope that this contribution encourages development of more sophisticated and realistic birth-death models in applied work, since researchers can now estimate parameters using more complicated rate structures, even when the data are observed at discrete times.

%%%%%%%%%%%%%%%%%%%%%%%%%%%%%%%%%%%%%%%%%%%%%%%%%%%%%%%%%%%
\section*{Software}

A software implementation of the EM algorithms for general BDPs used in this paper is currently available from FWC (by request through the Editor to maintain reviewer anonymity) and will be deposited in \citet{CRAN2011} before publication.

%%%%%%%%%%%%%%%%%%%%%%%%%%%%%%%%%%%%%%%%%%%%%%%%%%%%%%%%%%%
\section*{Acknowledgements}

We are grateful to Kenneth Lange, Hua Zhou, and Gabriela Cybis for helpful comments.  This work was supported by NIH grants R01 GM086887, HG006139, T32GM008185, and NSF grant DMS-0856099. 

%%%%%%%%%%%%%%%%%%%%%%%%%%%%%%%%%%%%%%%%%%%%%%%%%%

% BibTeX 
\bibliographystyle{spbasic}
\bibliography{fcrawford}

\end{document}